\documentclass[aps,prc,twocolumn,10pt,superscriptaddress]{revtex4-1}
 \usepackage{color}
 \usepackage{amsmath}
 \usepackage{graphicx}
 \usepackage{subfigure}
 \usepackage{multirow}
 \usepackage{mathrsfs}
 \usepackage{xcolor}
 \usepackage[colorlinks=true, linkcolor=black, citecolor=blue, urlcolor=blue]{hyperref}
\begin{document}

\title{Alpha-cluster Formation and Decay: Four-Body Correlation and Configuration Mixing}

\author{Yi Wu}
\affiliation{School of Physics, Nanjing University, Nanjing 210093, China}
\author{Chang Xu}
\affiliation{School of Physics, Nanjing University, Nanjing 210093, China}
\affiliation{China Institute of Atomic Energy, Beijing 102413, China}

\begin{abstract}
Alpha-decay provides an important probe of nuclear structure and the underlying nucleon-nucleon interaction, while its rigorous microscopic description from first principles remains challenging. For $\alpha$+magic core systems, a microscopic treatment of $\alpha$-cluster formation and decay has been achieved by considering both the four-body correlation on top of the core and Pauli blocking. Extending this microscopic description to open-shell nuclei should account for the configuration mixing caused by the residual interactions between valence nucleons near the Fermi surface. In this work, we improve the quartetting wave function approach (QWFA) by incorporating the pairing-induced configuration mixing using the particle-number-projected Bardeen-Cooper-Schrieffer (PBCS) method. We find that the pairing enhance the formation amplitudes of the $\alpha$-cluster in open-shell nuclei while the closed shells suppress the $\alpha$-clustering.

\end{abstract}

\maketitle
\section{Introduction}

As one of the spontaneous decay modes, $\alpha$-emission plays an important role in the study of heavy and superheavy nuclei and provides valuable information on $\alpha$-core interaction, nuclear shell effects and quantum tunneling in nuclear many-body systems~\cite{alpha1, alpha2, alpha3, alpha4, Superheavy1, Superheavy2, Wheeler, CQi1}. A microscopic understanding of the $\alpha$-cluster decay process is essential for improving $\alpha$-decay theories~\cite{Preformation, Cluster1, Cluster2, Cluster3, Bayesian, CFM, Mohr2006, Kelkar2007}. A key ingredient in the decay is the formation of the $\alpha$-cluster on top of the core, in which two protons and two neutrons develop strong correlations and form a compact cluster prior to emission. The theoretical description of $\alpha$-cluster formation can be formulated as an in-medium four-body problem, in which the $\alpha$-cluster is treated as a correlated $2p+2n$ quartet coupled to a core nucleus. To describe the in-medium motion of such a quartet, the center-of-mass (C.O.M.) motion of the quartet and its intrinsic degrees of freedom are treated simultaneously by solving the in-medium four-body equations~\cite{QWFA2014, QWFA2016, QWFA2017}. Owing to the Pauli blocking, the $\alpha$-cluster cannot remain bound deep inside the core nucleus and is expected to emerge only in the low-density surface region. Benchmark calculations have been performed for $\alpha$-emitters $^{104}$Te and $^{212}$Po~\cite{QWFA2020, QWFA2021} and compared with the available experimental data~\cite{Te1041, Te1042, Po212}. The prediction of an enhanced $\alpha$-cluster preformation factor in $^{104}$Te relative to $^{212}$Po is found to be consistent with recent experimental measurements~\cite{TePo}.

In extending the calculations to open-shell nuclei, residual interactions beyond the mean-field approximation should be taken into account as they can strongly mix different shell-model configurations. In particular, the pairing interaction between identical particles can generate collective pair motion and redistribute the occupation probabilities around the Fermi surface \cite{BCS1, BCS2, Pairing}. For example, it is well known that the two-nucleon transfer amplitude is significantly enhanced when the Bardeen-Cooper-Schrieffer (BCS) pairing is included in the theory of multi-nucleon transfer reactions~\cite{MNT1, Pairtransfer1, Pairtransfer2}. Such pairing-induced configuration mixing is expected to strengthen the coherent four-body correlations among two neutrons and two protons. The connection between pairing correlations and the $\alpha$-cluster formation amplitude has been studied using the BCS method~\cite{Delion1, Delion2, Delion3, CQi2}, in which the proton and neutron pairing correlations are treated separately, and the parent and daughter states are approximated by BCS quasiparticle vacua. The four nucleons forming the $\alpha$-particle are then described through coherent proton-pair and neutron-pair transfer amplitudes~\cite{Delion1, Delion2, Delion3, CQi2}. Pairing correlations thereby could induce the coherent mixing of many shell-model configurations and modify the $\alpha$-formation amplitude, whose contributions are weighted by the corresponding BCS occupation amplitudes. 

In the present work, we incorporate the pairing-induced configuration mixing of shell-model states in the quartetting wave function approach (QWFA) by using the particle-number-projected BCS (PBCS) method. The QWFA has been successfully applied to calculate the half-lives of $\alpha$-emitters in the vicinity of closed major shells 50, 82 and 126~\cite{QWFA2020, QWFA2021}. In open-shell regions, different configurations could mix strongly and the approaches such as PBCS method should be incorporated in order to achieve a better description of $\alpha$-cluster formation probability and decay width. We performed calculations for the $\alpha$-emitters of Po isotopes with mass numbers $A=188-208$. The quartet wave function is obtained by allowing multiple orbitals between the $N=82$ and $N=126$ shell closures to contribute collectively, rather than restricting it to a single shell-model state configuration. The correlation between the $\alpha$-cluster formation amplitudes and the strength of pairing correlations (pairing gap $\Delta$) is analyzed. By considering the pairing-induced configuration mixing, the calculated $\alpha$-decay half-lives for these isotopes are compared in detail with the experimental data.

This paper is organized as follows. In Sec.~II, we introduce the theoretical framework based on the QWFA combined with the PBCS method. In Sec.~III, we present the results for the Po isotopes and discuss the effects of pairing-induced configuration mixing on $\alpha$-cluster formation and $\alpha$-decay properties. Finally, Sec.~IV gives a summary.

\section{Theoretical Framework}

In the QWFA, the quartet wave function $\Psi_4({\bf R},{\bf s}_j)$ is decomposed into a C.O.M. part $\Phi({\bf R})$ and an intrinsic part $\varphi^{\mathrm{intr}}({\bf s}_j,{\bf R})$. The corresponding in-medium four-body equations for the quartet are
\begin{align}\label{C.O.M.}
&\hat{T}[\nabla_R]\,\Phi({\bf R})
-\frac{\hbar^2}{8m}\int {\rm d}^3s_j\bigg\{ \varphi^{\text{intr},*}({\bf s}_j,{\bf R})
[ \nabla_R^2 \varphi^{\text{intr}}({\bf s}_j,{\bf R})]
\nonumber\\
&\times\Phi({\bf R})\bigg\}+\int {\rm d}^3R'\,W({\bf R},{\bf R}')  \Phi({\bf R}')=E_4\,\Phi({\bf R})\, ,
\end{align}
\begin{align}\label{intr}
    &\hat{T}[\nabla_{s_j}]\,{\varphi}^{\text{intr}}({\bf s}_j,{\bf R})+\int{\rm d}^3R'{\rm d}^3s_j'\,\bigg\{V_4^{\text{intr}}({\bf s}_j,{\bf R};{\bf s}_j',{\bf R}')\,\nonumber\\
    &\times\frac{\Phi({{\bf R}'})}{\Phi({\bf R})}{\varphi}^{\text{intr}}({\bf s}_j,{\bf R}')\bigg\}=W^{\text{intr}}({\bf R})\,{\varphi}^{\text{intr}}({\bf s}_j,{\bf R})\,,
\end{align}
where $\hat{T}[\nabla_{R(s_j)}]$ is the kinetic energy operator for the C.O.M. (intrinsic) motion. As seen from Eqs.~(\ref{C.O.M.}) and (\ref{intr}), the C.O.M. part $\Phi({\bf R})$ and the intrinsic part $\varphi^{\mathrm{intr}}({\bf s}_j,{\bf R})$ are coupled in a nontrivial way. With the local density approximation, the effective C.O.M. potential can be written as
\begin{equation}\label{effW}
    \begin{aligned}
        W({\bf R})=\begin{cases}
        W_{\text{In}}({\bf R})\,, \,R<R_{\text{c}} \,,\\
        W_{\text{DF}}({\bf R})+W_{\text{PB}}({\bf R})-28.3\,\text{MeV} \,,\,R> R_{\text{c}}\,,
        \end{cases}
    \end{aligned}
\end{equation}
where the potential is defined piecewise with respect to the critical radius $R_{c}$, which corresponds to the critical density $\rho_c$. In the surface region ($R>R_{c}$), where the $\alpha$-cluster can be formed, $W({\bf R})$ consists of the $\alpha$-core interaction $W_{\text{DF}}({\bf R})$ obtained from the double-folding procedure and the isospin-dependent Pauli blocking term $W_{\text{PB}}({\bf R})$~\cite{Isospin}. In the interior region ($R<R_{c}$), the potential $W_{\text{In}}({\bf R})$ is determined by the C.O.M. density $\rho_4(R)$ of the valence nucleons. To obtain this density, one first integrates over the relative coordinates and then performs the convolution in momentum space through a Fourier transformation
\begin{align}\label{psiab}
    \psi_{ab}({\bf p})&=\frac{1}{(2\pi)^6}\int{\rm d}^3r_a{\rm d}^3r_b
    \left|\Psi_{jj}^0(a,b)\right|^2{\rm e}^{-{\rm i}{\bf p}\cdot{\bf r}_a}{\rm e}^{-{\rm i}{\bf p}\cdot{\bf r}_b}\,,
\end{align}
where $\Psi_{jj}^0(a,b)$ denotes a two-body wave function in which two nucleons occupying the same shell-model orbital $j$ are coupled to total angular momentum zero. The coordinates $(a,b)=(1,2)$ refer to the neutron pair, while $(a,b)=(3,4)$ correspond to the proton pair. In terms of the single-particle wave function $\phi_{jm}$, $\Psi_{jj}^0(a,b)$ can be written as
\begin{align}\label{Psiab}
    \Psi_{jj}^0(a,b)
    =\sum_{m}\langle jmj-m|00\rangle\phi_{jm}(a)\phi_{j-m}(b)\,.
\end{align}
The corresponding quartet C.O.M. density obtained from the the quartet wave function $\Psi_4$ of the valence nucleons is then given by
\begin{align}\label{rho}
    \rho_4(R)&=\int{\rm d}^3s_j|\Psi_4({\bf R},{\bf s}_j)|^2\nonumber\\
    &=2^6(2\pi)^9\int{\rm d}^3p\,\psi_{12}({\bf p})\psi_{34}({\bf p}){\rm e}^{4{\rm i}{\bf p}\cdot{\bf R}}\,.
\end{align} 

The single shell-model state configuration approximation used above is expected to work best near major shell closures (such as $Z=82$ and $N=126$ in nuclei like $^{210}$Pb, $^{210}$Po, and $^{212}$Po), where the large shell gaps reduce the single-particle level density around the Fermi surface and thus suppress pairing correlations~\cite{Shellgap}. In extending to open-shell nuclei, pairing mixes nearby orbitals and allows several shell-model configurations to contribute coherently to the formation of the $\alpha$-like quartet. For this reason, we focus on nucleon pairing correlations and describe the valence nucleons by the Hamiltonian
\begin{align}\label{meanfield}
\widehat{H}(\lambda)&=\sum_{\nu>0}(\varepsilon_{\nu}-\lambda)\big(\hat{c}_{\nu}^{\dagger}\hat{c}_{\nu}+\hat{c}_{\bar{\nu}}^{\dagger}\hat{c}_{\bar{\nu}}\big)-G\sum_{\nu,\mu>0}\hat{c}_{\nu}^{\dagger}\hat{c}_{\bar{\nu}}^{\dagger}\hat{c}_{\bar{\mu}}\hat{c}_{\mu}\,,
\end{align}
where $G$ is the strength of the pairing interaction, and the creation (annihilation) operator $\hat{c}_{\nu}^{\dagger}$ ($\hat{c}_{\nu}$) creates (annihilates) a particle in the single-particle state $|\nu=jm\rangle$, and $\bar{\nu}$ denotes the time-reversed state labeled by $\nu$. The pairing strength $G$ is determined by reproducing the empirical pairing gap $\Delta$~\cite{AME, Gap}.

The pairing Hamiltonian (Eq.~(\ref{meanfield})) is treated separately for the parent (P) and daughter (D) nuclei within the BCS approximation, yielding their corresponding ground states. Since the BCS ground state does not conserve particle number exactly, the projected ground states are constructed by applying the particle-number projection operator $\widehat{P}^{N_0}$,
\begin{align}\label{groundstate}
\begin{cases}
|\Psi_{\text{P}}\rangle=\displaystyle\frac{\widehat{P}^{N_0}}{\mathscr{N}_{\text{P}}}|\text{BCS}_\text{P}\rangle
\\
\\
|\Psi_{\text{D}}\rangle=\displaystyle\frac{\widehat{P}^{N_0-2}}{\mathscr{N}_{\text{D}}}|\text{BCS}_\text{D}\rangle
\end{cases}\,,
\end{align}
here
\begin{align}\label{PN}
\widehat{P}^{N_0}=\frac{1}{2\pi}\int_0^{2\pi}{\rm e}^{{\rm i}\varphi(\widehat{N}-N_0)}{\rm d}\varphi\,.
\end{align}
$\mathscr{N}_{\text{P}}$ and $\mathscr{N}_{\text{D}}$ are the normalization factors for the projected parent and daughter states.

The corresponding quartet wave function can then be written as $\Psi_4=\frac{1}{2}\langle\Psi_{\text{D}}|\hat{\psi}({\bf r}_4)\hat{\psi}({\bf r}_3)\hat{\psi}({\bf r}_2)\hat{\psi}({\bf r}_1)|\Psi_{\text{P}}\rangle=\Psi_{12}\Psi_{34}$, and the nucleon pair transfer amplitude $\Psi_{ab}$ is
\begin{widetext}
\begin{align}\label{psip}
    \Psi_{ab}&=\frac{1}{\sqrt{2}}\,_{\nu(\pi)}\langle\Psi_{\text{D}}|\hat{\psi}({\bf r}_b)\hat{\psi}({\bf r}_a)|\Psi_{\text{P}}\rangle_{\nu(\pi)}\nonumber\\
     &=\frac{1}{\mathscr{N}_{\text{P}}\mathscr{N}_{\text{D}}}\frac{1}{2\pi{\rm i}}\oint\frac{{\rm d}z}{z^{\frac{N_0-2}{2}+1}}\sum_{j}\sqrt{j+\frac{1}{2}}u^{(\text{D})}_{j}v^{(\text{P})}_{j}\Psi_{jj}^0(a,b)\left[u^{(\text{D})}_{j}u^{(\text{P})}_{j}+zv^{(\text{D})}_{j}v^{(\text{P})}_{j}\right]^{\Omega_j-1}\nonumber\\
     &\times\prod_{k\neq j}\left[u^{(\text{D})}_{k}u^{(\text{P})}_{k}+zv^{(\text{D})}_{k}v^{(\text{P})}_{k}\right]^{\Omega_k}\nonumber\\
     &=\sum_{j}\sqrt{j+\frac{1}{2}}\,u^{(\text{D})}_{j}v^{(\text{P})}_{j}\Psi_{jj}^0(a,b)\frac{\mathscr{I}_{j}}{\mathscr{N}_{\text{P}}\mathscr{N}_{\text{D}}}\,,
\end{align}
\end{widetext}
where $\Omega_j = j + \frac{1}{2}$ denotes the pair degeneracy of the shell-model orbital with angular momentum $j$ and $\nu(\pi)$ labels the neutron (proton) sector. The $v_{\nu}^{(\text{P}(\text{D}))}$ and $u_{\nu}^{(\text{P}(\text{D}))}$ are the usual BCS occupation and vacancy amplitudes of parent (daughter) nuclei, respectively. The transfer amplitude is a coherent superposition of the two-particle components $\Psi_{jj}^{0}(a,b)$ associated with different shell-model orbitals $j$, with weights determined by the occupation amplitudes and the number-projection integrals. In the limit of vanishing pairing ($G=0$), this coherence disappears, and the pair transfer amplitude essentially reduces to the single dominant configuration $\Psi_{jj}^0(a,b)$ with $u_j^{(\text{D})}v_j^{(\text{P})}\neq0$.

Once the configuration-mixed quartet wave function $\Psi_4$ is obtained, the internal part of the C.O.M. potential $W_{\text{In}}(R)$ can be calculated from the quartet density $\rho_4(R)$~\cite{QWFA2020, QWFA20172}
\begin{align}\label{Internal}
    W_{\text{In}}(R)-E_4&=\frac{\hbar^2}{8m}\frac{\rho'_4(R)}{R\,\rho_4(R)}-\frac{\hbar^2}{32m}\frac{\rho'_4(R)^2}{\rho_4(R)^2}\nonumber\\
    &+\frac{\hbar^2}{16m}\frac{\rho_4''(R)}{\rho_4(R)}\,,
\end{align}
where $E_4$ denotes the energy of the quartet. In this way, pairing-induced configuration mixing modifies the quartet density $\rho_4(R)$, which in turn changes the internal potential $W_{\text{In}}(R)$ and eventually the C.O.M. wave function relevant for the $\alpha$-cluster formation.

\section{Results}

\begin{figure}[htbp]
\begin{minipage}{1\linewidth}
\centerline{\includegraphics[width=1\textwidth]{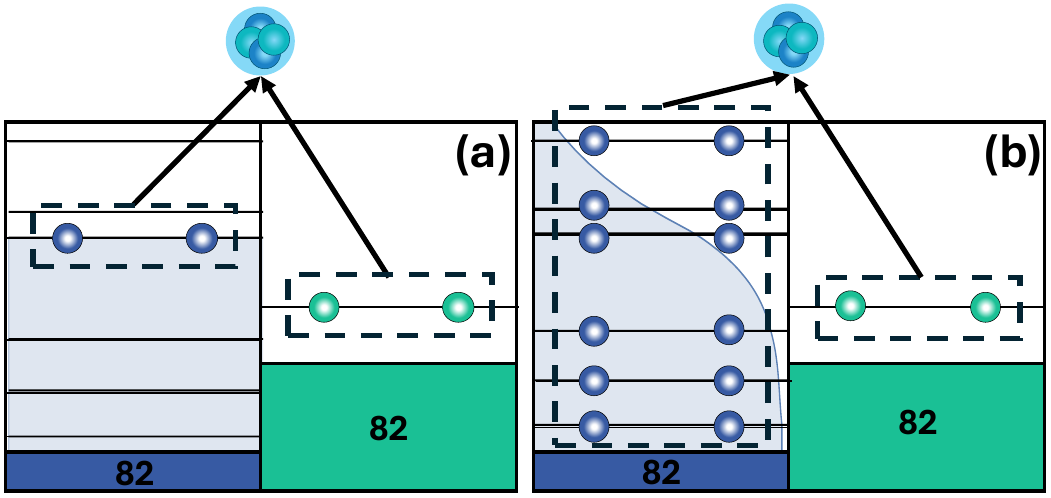}} 
\end{minipage}
\caption{Schematic illustration of valence nucleons and $\alpha$-cluster formation in $^{200}$Po. Panel (a): The scenario in the absence of configuration mixing. The orbitals below the Fermi surface are fully occupied. Panel (b): The configuration mixing is included and orbitals become partially occupied.}
\label{Energylevel}
\end{figure}

The pair transfer amplitude may receive contributions from many shell-model orbitals around the Fermi surface once pairing is taken into account (see Eq.~(\ref{psip})). Therefore, the first step of the analysis is to identify the valence space considered here. A useful guide is provided by the empirical $N_pN_n$ scheme~\cite{NnNp1, NnNp2}, which shows that, in the heavy mass region, the structural evolution is largely determined by the valence nucleons outside the nearest shell closure. In the Po isotopic chain, the proton number is fixed at $Z=84$, with only two valence protons lying outside the $Z=82$ shell closure, so that the proton-pairing-induced configuration mixing is expected to be limited, whereas the much larger neutron valence space between $N=82$ and $N=126$ makes the neutron sector the dominant source of configuration mixing along the isotopic chain. On this basis, the Po isotopes offer a suitable framework for studying how the neutron-pairing-induced configuration mixing in the relevant valence space influences the $\alpha$-cluster formation. We therefore consider only the neutron-pairing effects in the following analysis. Accordingly, the neutron valence space considered in the present work consists of the orbitals between the $N=82$ and $N=126$ shell closures, namely $1f_{7/2}$, $0h_{9/2}$, $0i_{13/2}$, $2p_{3/2}$, $1f_{5/2}$, and $2p_{1/2}$, and only the protons from $0h_{9/2}$ orbital are taken into account, as schematically illustrated in Fig.~\ref{Energylevel}.

\begin{figure}[htpb]
\begin{minipage}{1\linewidth}
\centerline{\includegraphics[width=1\textwidth]{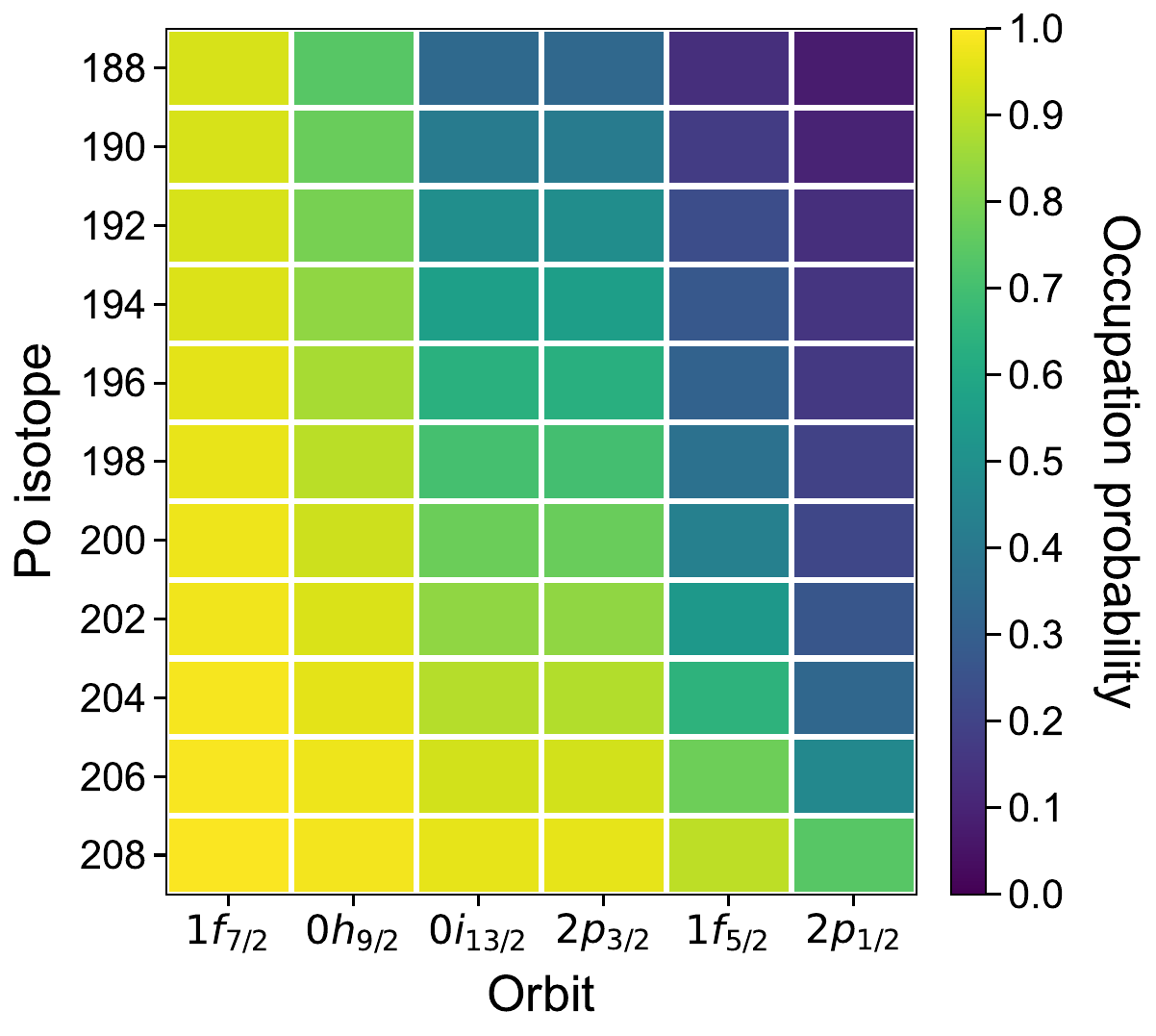}}
\end{minipage}
\caption{The calculated occupation probabilities $v^2_{\nu}$ of neutron shell-model states for $^{188-208}$Po isotopes. The lower-lying orbitals are nearly fully occupied ($v^2_\nu \approx 1$), whereas pairing correlations lead to fractional occupations of the higher-lying orbitals around the Fermi surface.}
\label{Occupancy}
\end{figure}

\begin{figure}[htpb]
\begin{minipage}{1\linewidth}
\centerline{\includegraphics[width=1\textwidth]{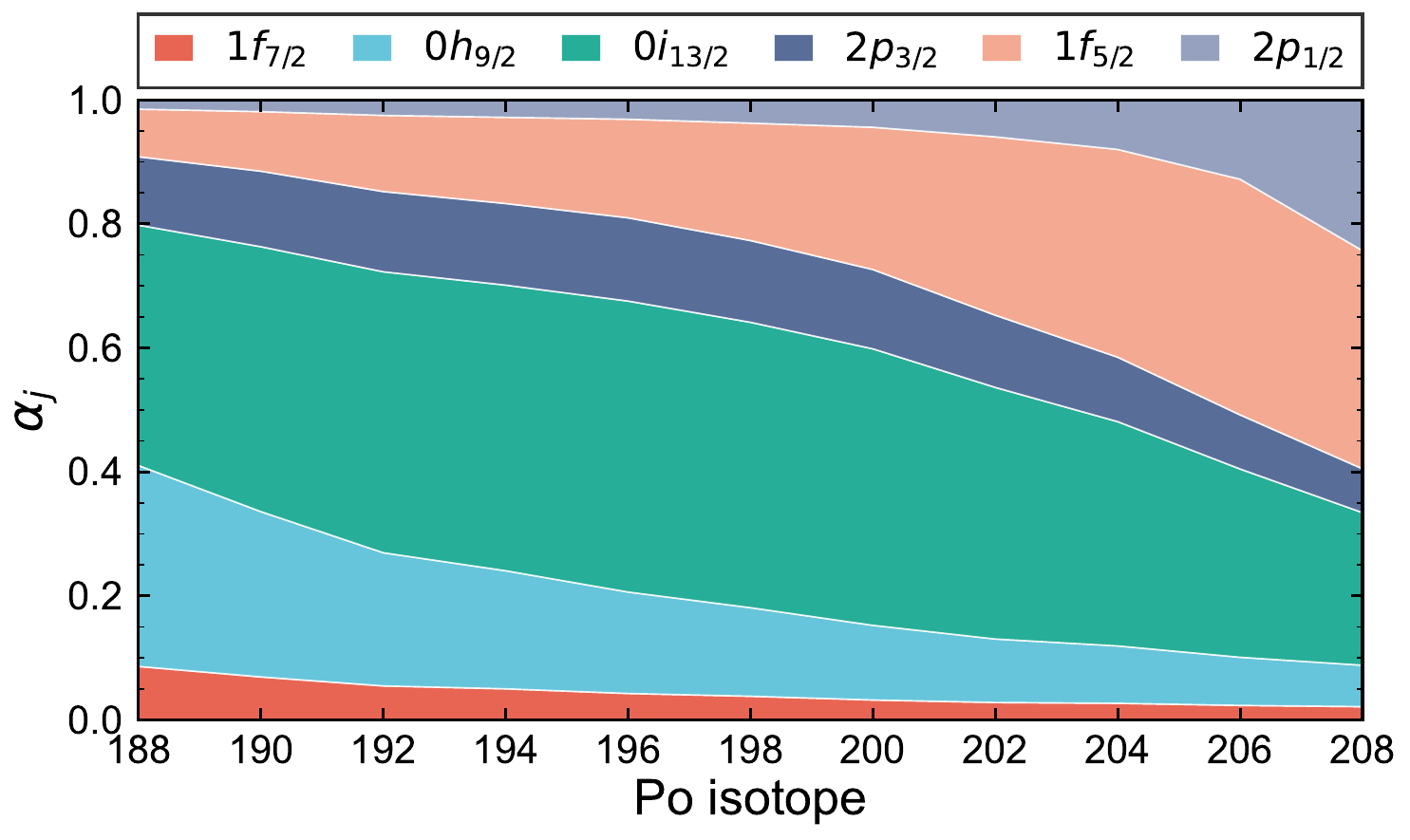}}
\end{minipage}
\caption{Fractional weights ($\alpha_j$) of individual orbitals in the pair transfer amplitude for different isotopes. The weights are normalized such that $\sum_j \alpha_j = 1$. The dominant contributions come from the orbitals near the Fermi surface.}
\label{alphastacked}
\end{figure}

In this valence space, the occupation probabilities $v_{\nu}^2$ for the Po isotopes considered here are shown in Fig.~\ref{Occupancy}. As the neutron number increases, the pairing correlations lead to partial occupancies in orbitals above and below the Fermi level. The contribution of a given orbital to the quartet wave function is governed by the corresponding occupation and vacancy amplitudes together with the projection factor, as can be seen in Eq.~(\ref{psip}), and is therefore expected to be largest for orbitals closest to the Fermi surface. To make this point explicit, we define the fractional weight $\alpha_j$ of each orbital $j$ in the neutron-pair transfer amplitude as
\begin{align}\label{alpha}
    \alpha_{j}=\frac{\left|\sqrt{j+\frac{1}{2}}\,u^{(\text{D})}_{j}v^{(\text{P})}_{j}\frac{\mathscr{I}_{j}}{\mathscr{N}_{\text{P}}\mathscr{N}_{\text{D}}}\right|^2}{\displaystyle\sum_{j}\left|\sqrt{j+\frac{1}{2}}\,u^{(\text{D})}_{j}v^{(\text{P})}_{j}\frac{\mathscr{I}_{j}}{\mathscr{N}_{\text{P}}\mathscr{N}_{\text{D}}}\right|^2}\,,
\end{align}
Fig.~\ref{alphastacked} shows that the dominant contributions come from the orbital nearest the Fermi surface, while neighboring orbitals remain non-negligible because of pairing-induced configuration mixing.

\begin{figure}[htbp]
\begin{minipage}{1\linewidth}
\centerline{\includegraphics[width=1\textwidth]{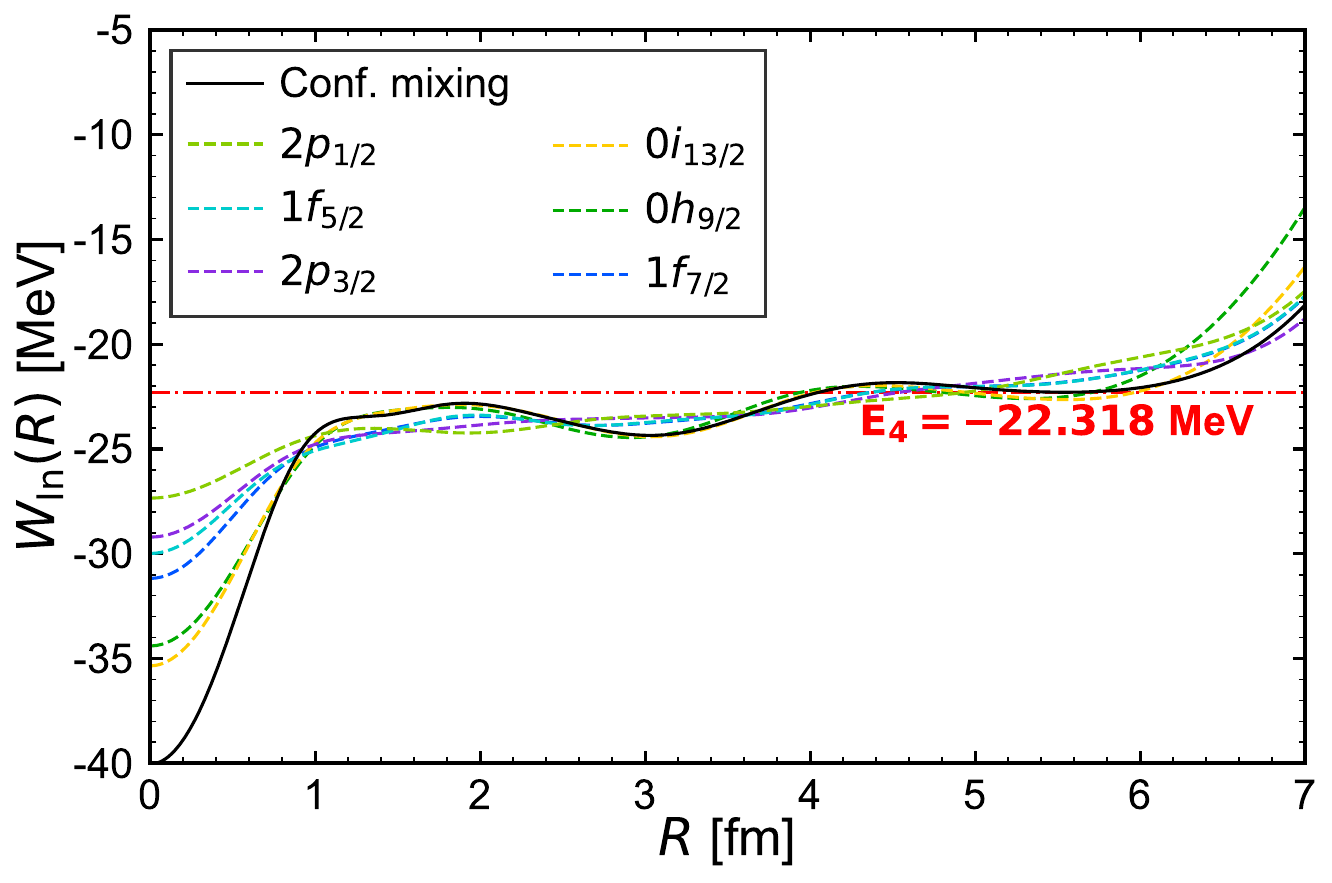}}
\end{minipage}
\caption{The C.O.M. effective potential, $W_{\text{In}}(R)$, for $^{200}$Po. The dashed curves show the $W_{\text{In}}(R)$ obtained from different neutron orbitals coupled to a fixed proton orbital. The horizontal red dash-dotted line indicates the quartet energy, $E_4 = B_{\text{D}}-B_{\text{P}} = -22.318$ MeV, where $B_{\text{P}}$ and $B_{\text{D}}$ are the binding energies of the parent and daughter nuclei~\cite{AME}, respectively.}
\label{Orbit}
\end{figure}

Once the pairing-induced configuration mixing is included for the Po isotopes, the quartet wave function $\Psi_4$ becomes a coherent superposition of different shell-model configurations, which can be written as
\begin{align}\label{mixing}
\Psi_4&=C_1(\nu\,1f_{7/2})^2_{J_n=0}\otimes(\pi\,0h_{9/2})^2_{J_p=0}\nonumber\\
&+C_2(\nu\,0h_{9/2})^2_{J_n=0}\otimes(\pi\,0h_{9/2})^2_{J_p=0}\nonumber\\
&+C_3(\nu\,0i_{13/2})^2_{J_n=0}\otimes(\pi\,0h_{9/2})_{J_p=0}^2+...\,,
\end{align}
where $C_j=\sqrt{j+1/2}u_j^{(\mathrm{D})}v_j^{(\mathrm{P})}\frac{\mathscr{I}_j}{\mathscr{N}_{\mathrm{P}}\mathscr{N}_{\mathrm{D}}}$ is obtained from Eq.~(\ref{psip}). Its norm $\mathscr{N}_4=\displaystyle\int{\rm d}^3r_1{\rm d}^3r_2{\rm d}^3r_3{\rm d}^3r_4\left|\Psi_{4}\right|^2$ characterizes the strength of the quartet wave function relative to the no-mixing case ($\mathscr{N}_4^{(0)}=1$). The superposition of the different shell-model configurations modifies the spatial distribution $\rho_4(R)$ of the quartet as well as the internal part of the quartet C.O.M. potential $W_{\text{In}}(R)$. Fig.~\ref{Orbit} displays $W_{\text{In}}(R)$ obtained from configuration mixing, in comparison with those obtained from individual configurations. It can be seen that $W_{\text{In}}(R)$ reflects the radial characteristics of the valence-nucleon shell-model states involved. The behavior of $W_{\text{In}}(R)$ depends on the radial distributions of the single-particle wave function. Stronger confinement to the nuclear interior generally leads to a steeper increase with $R$. In the configuration-mixed case, interference between components modifies the radial structure. The resulting $W_{\text{In}}(R)$ therefore differs from that obtained from each individual configuration. By substituting $W_{\text{In}}(R)$ into Eq.~(\ref{C.O.M.}), the corresponding C.O.M. wave function, $\sqrt{\mathscr{N}_4}\Phi({\bf R})$, can be obtained, where $\Phi({\bf R})$ is the normalized C.O.M. wave function and the prefactor $\sqrt{\mathscr{N}_4}$ characterizes the strength of the quartet wave function.

\begin{figure}[htbp]
\begin{minipage}{1\linewidth}
\centerline{\includegraphics[width=1\textwidth]{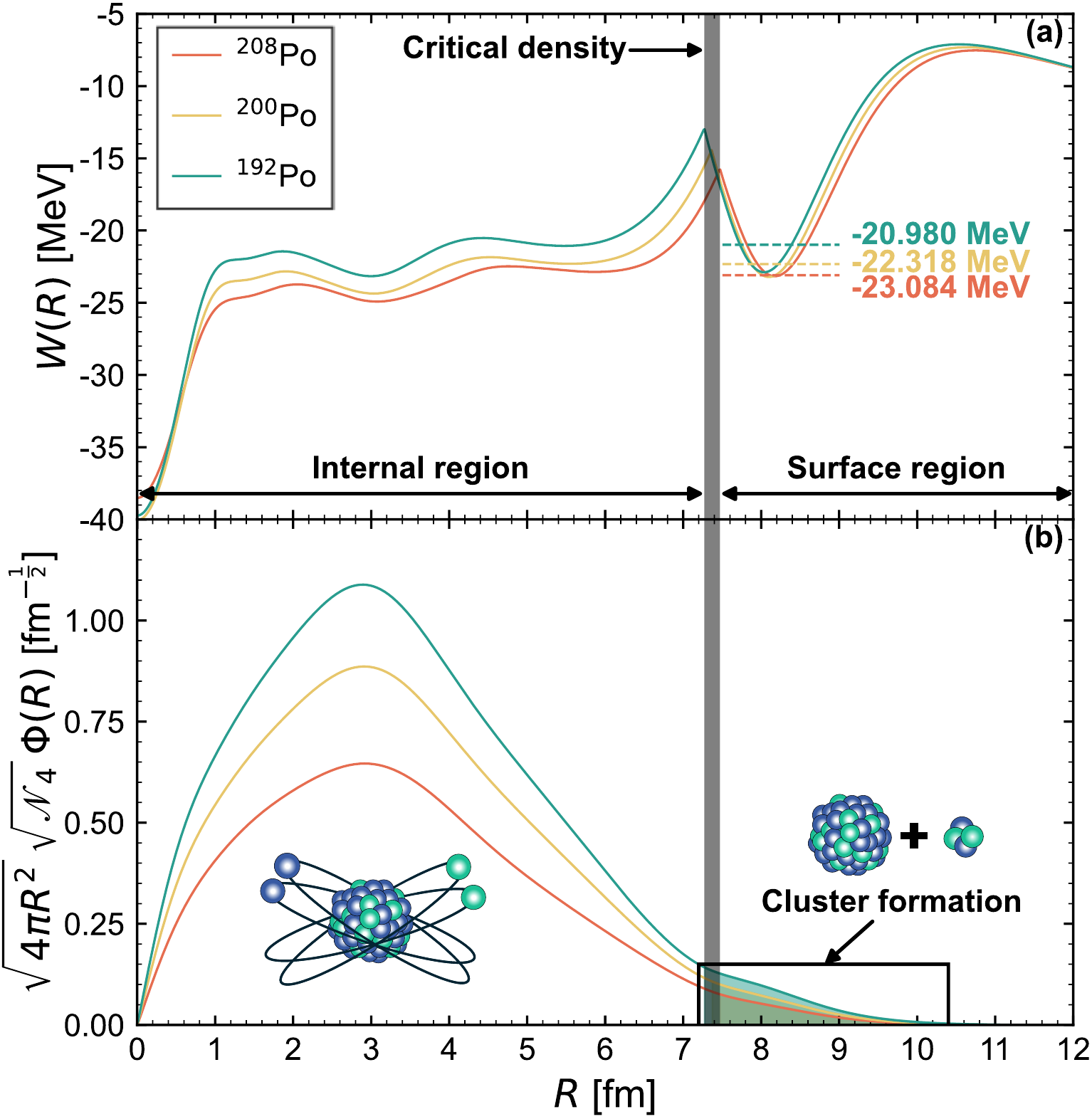}}
\end{minipage}
\caption{(a) The C.O.M. effective potentials and (b) the corresponding C.O.M. wave functions for $^{192}$Po, $^{200}$Po, and $^{208}$Po.}
\label{QWFA}
\end{figure}

The $\alpha$-cluster formation factor can be obtained by integrating the C.O.M. probability density from the critical radius $R_c$ to infinity, since a bound $\alpha$-cluster can survive only in the nuclear surface region owing to the strong Pauli-blocking effect in the nuclear interior. It is therefore defined as
\begin{align}\label{palpha}
P_{\alpha}
=\mathscr{N}_4\int {\rm d}^3R\,|\Phi({\bf R})|^2\Theta(R-R_c),
\end{align}
where $\Theta(R-R_c)$ is the Heaviside step function, which restricts the integration to the region $R\geq R_c$. In the absence of configuration mixing, $\mathscr{N}_4\rightarrow1$, and $P_{\alpha}$ automatically reduces to the no-mixing case $P_{\alpha}
=\displaystyle\int {\rm d}^3R\,|\Phi({\bf R})|^2\Theta(R-R_c)$~\cite{QWFA2020,QWFA2021}. The C.O.M. wave function $\sqrt{\mathscr{N}_4}\Phi({\bf R})$ and the corresponding C.O.M. effective potentials for $^{192}$Po, $^{200}$Po and $^{208}$Po are shown in Fig.~\ref{QWFA}. As shown in Fig.~\ref{QWFA}(a), a pronounced pocket develops in the nuclear surface region beyond the critical radius $R_c$, where the Pauli-blocking effect is sufficiently weakened and a bound $\alpha$-cluster can survive. Correspondingly, the shaded part of the C.O.M. wave function in Fig.~\ref{QWFA}(b) highlights the component extending into this surface pocket, which contributes dominantly to the formation of the $\alpha$-cluster.

\begin{figure}[htbp]
\begin{minipage}{1\linewidth}
\centerline{\includegraphics[width=1\textwidth]{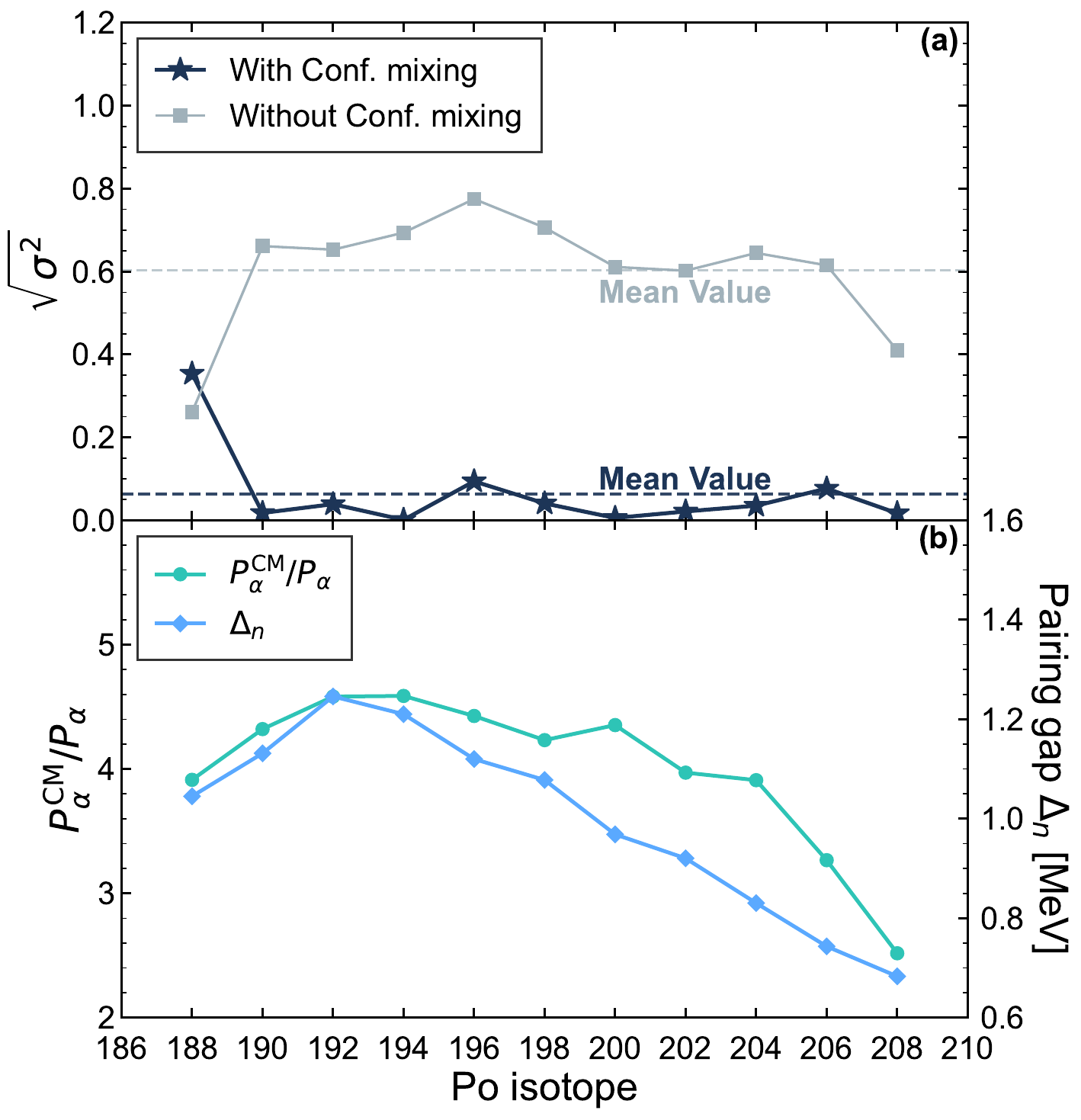}}
\end{minipage}
\caption{(a) The dark blue curve shows the absolute logarithmic deviation of the half-lives, $\sqrt{{\sigma^2}}$, with configuration mixing, while the gray curve without configuration mixing. (b) The green curve denotes the enhancement ratio of the formation factor, $P_{\alpha}^{\mathrm{CM}}/P_{\alpha}$, and the blue curve denotes the neutron pairing gap $\Delta_n$.}
\label{Result}
\end{figure}

The $\alpha$-decay half-life $T_{1/2}^{\mathrm{Cal.}}$ is then calculated within the two-potential approach (TPA)~\cite{TPA}. For comparison with experiment, we define $\sigma=\text{log}_{10}T_{1/2}^{\text{Exp.}}-\text{log}_{10}T_{1/2}^{\text{Cal.}}$~\cite{NNDC}. Fig.~\ref{Result} summarizes the resulting absolute logarithmic deviation $\sqrt{\sigma^2}$ obtained with and without configuration mixing, together with the enhancement of the formation factor $P_{\alpha}^{\mathrm{CM}}/P_{\alpha}$ and the neutron pairing gap $\Delta_n$. Here, $P_{\alpha}^{\mathrm{CM}}$ and $P_{\alpha}$ denote the formation factors calculated with and without configuration mixing, respectively.

As shown in Fig.~\ref{Result}, the inclusion of configuration mixing improves the overall agreement between the calculated and experimental half-lives. This improvement can be traced to the fact that the pairing correlations enhance the coherence of the quartet wave function, and thereby increase the $\alpha$-cluster formation factor. The gradual reduction of the enhancement of the formation factor $P_{\alpha}^{\mathrm{CM}}/P_{\alpha}$ toward the $N=126$ shell closure can be understood in terms of the competition between pairing coherence and shell stabilization. As the neutron Fermi surface approaches the large shell gap, the level density around the Fermi surface decreases and the available phase space for pair scattering becomes restricted. The coherent admixture of neighboring orbitals is then suppressed, so that the quartet wave function becomes closer to the no-mixing limit. From this viewpoint, the behavior near $N=126$ is not simply a reduction in pairing correlations, but a transition from a correlation-dominated regime in open-shell nuclei to a more single-particle-like regime near the closed shell. 

Starting from the vicinity of the $N=126$ shell closure and moving toward the open-shell region, neutron pairing correlations become stronger and induce increasingly coherent mixing among the valence-neutron configurations, leading to a corresponding increase in the predicted $\alpha$-cluster formation factor. This systematic behavior supports the important role of pairing-induced configuration mixing in enhancing $\alpha$-cluster formation away from the shell closure. 

However, this purely pairing-induced picture becomes insufficient toward the neutron mid-shell around $N=104$, where the calculated enhancement tends to deviate from the experimental trend, most notably for $^{188}$Po in Fig.~\ref{Result}. Additional collective effects, such as deformation and shape coexistence, may become important in this region. Low-lying deformed intruder configurations compete with the normal configuration and may contribute to the ground-state wave functions of the parent Po nuclei. By contrast, the ground states of the daughter Pb nuclei remain dominated by normal, near-spherical configurations owing to the $Z=82$ shell closure~\cite{FRDM}. This parent–daughter configuration mismatch may hinder $\alpha$-cluster formation and suppress the formation factor~\cite{Intruder1, Intruder2, Intruder3}. The anomalous behavior of the lighter Po isotopes therefore suggests that a single PBCS ground state is inadequate. Shape coexistence and the mixing of normal and intruder configurations should also be taken into account.

\section{Summary}

In this work, we present a microscopic description of the $\alpha$-cluster formation and decay by considering the pairing-induced configuration mixing in the QWFA. The pairing correlations are treated within the particle-number-projected BCS (PBCS) framework, which allows the contributions from the shell-model orbitals around the Fermi surface to be coherently redistributed within the valence space between the $N=82$ and $N=126$ shell closures. We show that, in open-shell nuclei, the pairing-induced configuration mixing enhances the quartet wave function by allowing multiple orbitals to contribute collectively, rather than restricting the quartet wave function to a single shell-model state configuration. In contrast, near the $N=126$ shell closure, the configuration mixing effects induced by pairing are suppressed by the large shell energy gap. For the neutron-deficient nuclides closer to the neutron mid-shell $N=104$, the collective correlations, especially nuclear deformation, may lead to a configuration mismatch between the parent and daughter states and a suppression of the $\alpha$-cluster formation factor. We also show that the overall agreement between the calculated and experimental $\alpha$-decay half-lives for the $^{188-208}$Po isotopes is improved by considering the configuration mixing of different shell-model states. 

One direction for future work is to extend the present framework beyond a single PBCS configuration by including the mixing of coexisting shape configurations, thereby allowing possible shape mismatches between the parent and daughter nuclei to be treated explicitly. Such an extension would provide a more consistent description of the structural differences between the parent and daughter nuclei and help clarify how pairing correlations and shape coexistence jointly influence the $\alpha$-cluster formation.

\section{Acknowledgments}

This work is supported by the National Natural Science Foundation of China (Grant No. 12275129) and by the Fundamental Research Funds for the Central Universities (Grant No. 020414380257).

\end{document}